# Pressure induced structural transitions and metallization in $Ag_2Te$

Zhao Zhao[1], Shibing Wang[2,3], Haijun Zhang[1], Wendy L. Mao[2,4]

[1] *Department of Physics, Stanford University, Stanford, CA 94305, USA*

[2] *Department of Geological and Environmental Sciences, Stanford University, Stanford, CA 94305, USA*

[3] *Stanford Synchrotron Radiation Light Source, SLAC National Accelerator Laboratory, Menlo Park, CA 94025, USA*

[4] *Photon Science, SLAC National Accelerator Laboratory, Menlo Park, CA 94025, USA*

**Abstract:**

High pressure *in-situ* synchrotron X-ray diffraction experiments were performed on $Ag_2Te$ up to 42.6 GPa at room temperature and four phases were identified. Phase I ($\beta – Ag_2Te$) transformed into phase II at 2.4 GPa, and phase III and phase IV emerged at 2.8 GPa and 12.8 GPa respectively. Combined with first-principles calculations, we solved the phase II and phase III crystal structures, and determined the compressional behavior of phase III. Electronic band structure calculations show that the insulating phase I with a narrow band gap first transforms into semi-metallic phase II with the perseverance of topologically non-trivial nature, and then to bulk metallic phase III. Density of States (DOS) calculations indicate the contrasting transport behavior for $Ag_{2-\delta}Te$ and $Ag_{2+\delta}Te$ under compression. Our results highlight pressure's dramatic role in tuning $Ag_2Te$'s electronic band structure, and its novel electrical and magneto transport behaviors.

Silver telluride, $Ag_2Te$, is a superionic conductor [1,2], a novel material composed of non-magnetic atoms showing large linear magnetoresistance (LMR) [3], and predicted to be a binary 3D topological insulator with a highly anisotropic single Dirac cone on the surface [4]. At room temperature and ambient pressure, bulk $Ag_2Te$ is a narrow band gap insulator with a monoclinic crystal structure (Space group $P2_1/c$, Z = 4) called $\beta$ – $Ag_2Te$. Its strongly distorted antifluorite structure has a triple layered Te (Ag) – Ag – Te (Ag) stacking structure where the Te atoms occupy a distorted face-centered cubic (FCC) lattice with Ag atoms inserted in the interstitials [5,6] (Fig. 1(b)). When heated above 417 K, it transforms into superionic $\alpha$ – $Ag_2Te$, which has a FCC structure (Space group $F m \bar{3} m$, Z = 4) [1,2].

Unusually large LMR was observed in $Ag_2Te$ in magnetic fields up to 55 kOe, from 5K to room temperature [3], making it a promising material for industrial applications, such as magnetic field sensors fabrication [7]. To explain the novel LMR, classical and quantum explanations have been proposed. The classical solution emphasized the inhomogeneous distribution of Ag ions and large spatial fluctuations in the conductivity of a 2D system where the gap goes to zero [8], also supported by transport measurements [9,10]. While the quantum solution assumed a gapless spectrum with linear momentum dependence between the valence and conduction bands and strong disorder in the $Ag_2Te$ system [11,12]. Interestingly, $Ag_2Te$ was recently predicted to be a 3D topological insulator [4], which is an area that attracts increasing research interest [13–16]. Experimental evidence of the pronounced Aharanov-Bohm oscillation observed in single crystalline $\beta$-$Ag_2Te$ nanowires and nanoribbons [6,17] similar to the case of nano-sized $Bi_2Se_3$ and Bi$_2$Te$_3$ [18,19] suggests the existence of metallic surface states, which offers a possible origin of the LMR [4]. In addition, the Dirac cone of *β*-Ag$_2$Te is assumed to be highly anisotropic due to its monoclinic symmetry [4], differing from previous well studied 3D topological insulators like Bi$_2$Se$_3$ and Bi$_2$Te$_3$ that have an isotropic Dirac cone [20–24]. Novel physics like spin conduction are expected for a topological insulator with a highly anisotropic Dirac cone [25,26].

Pressure is a powerful tool for tuning materials' properties, because it can induce dramatic changes in the interatomic distances and atomic arrangements. A previous electrical and magneto transport study on Ag$_2$Te showed that pressure has significant effects on the electrical resistivity and MR behaviors, and can modify the electronic band structure [27]. Despite these interesting high pressure phenomena, only few high pressure structural studies on Ag$_2$Te have been reported, and the space group of the high pressure phases have not been determined [28]. To solve the Ag$_2$Te structures and explore their related electronic properties at high pressure, we performed *in-situ* synchrotron angle dispersive powder X-ray diffraction (XRD) experiments and first-principles calculations on Ag$_2$Te. In this paper, we report the new structural models of Ag$_2$Te at high pressure and their calculated electronic band structures.

In our experiments, high purity powder Ag$_{2+\delta}$Te (δ=0.03) was purchased from Sigma-Aldrich supplier. Symmetric diamond anvil cells with 300 microns culet size were used. Tungsten thin foils were used as the gasket and a 120um diameter sample chamber was drilled in the center. Ruby spheres were used for determining pressure. In two separate experiments, silicone oil was used as the pressure medium to study the low pressure region up to 4.8 GPa and neon gas was used as the pressure medium to maintain hydrostatic condition to reach 42.6 GPa.

Angle dispersive XRD experiments were performed at beamlines 16-IDB and 16-BMD of the Advanced Photon Source (APS), Argonne National Laboratory (ANL), and 12.2.2 of the Advanced Light Source (ALS), Lawrence Berkeley National Laboratory (LBNL). Jade was used for space group indexing [29], and the Rietveld fitting was performed using GSAS and EXPGUI package [30].

The Vienna *ab-initio* Simulation Package (VASP) [31,32] was employed for the crystal structure relaxations with the framework of Perdew-Burke-Ernzerhof (PBE) type [33] generalized gradient approximation (GGA) of density functional theory [34]. The projector augmented wave (PAW) [35] pseudo-potential was used for all the calculations. As the phase II structure is a modification of the phase I structure with the same space group, we only locally relaxed the crystal structure of phase II from our experiments. In order to double check the crystal structure of phase III from our experiments, we used the evolutionary method to search for the globally minimum enthalpy though USPEX [36–38]. The volume was fixed at the experimental value and the kinetic energy cutoff was fixed to 350eV and the resolution of $k$-mesh was taken from 0.2 to 0.08 ($2\pi\text{Å}^{-1}$) linearly for the calculations of USPEX. To calculate the band structures of all the phases, the kinetic energy cutoff was fixed to 450eV and spin-orbit coupling interaction was included through the non self-consistent calculation.

Representative XRD patterns are shown in Fig. 1(a). New peaks and those with significantly different intensity are marked to distinguish new phases. The lowest pressure diffraction measurement at 1.2 GPa confirms the ambient *β*-$Ag_2Te$ (phase I) monoclinic structure [5]. The measurement taken at 2.4 GPa shows a different pattern and indicates the appearance of phase II. Phase III emerges at 2.8 GPa and the transformation completes at 3.2 GPa. Phase III starts to transform into phase IV at 12.8 GPa and transforms completely at 17.2 GPa. Phase IV persists to the highest pressure measured. Under compression, the largest $d$ spacing peak corresponds to the stacking distance of triple Te(Ag) – Ag – Te(Ag) layered structure and shifts continuously to smaller $d$ spacings up to 42.6 GPa. Decompression experiments show that all the phase transitions are reversible.

Fig. 1(b) shows the structures of phases I, II, and III. Phase I has a monoclinic structure with 4 $Ag_2Te$ per unit cell and two different Ag sites [5]. We found that phase II has a monoclinic structure with space group *$P2_1/c$*, Z = 4, and phase III has an orthorhombic structure with space group *Aba2*, Z = 8. Representative Rietveld refinement results and profiles are presented in supplementary material [39]. From phase I to phase II, the strong decrease in the intensity of peaks like $(111)$ and $(21\bar{1})$ were observed without additional new peaks, thus phase II is assigned to the same monoclinic space group of phase I. In previous literature [28], the features (see the Supplementary Fig. 1(a)), which we identify as separate $(112)$ and $(31\bar{2})$ peaks, were considered as a single reflection. Our data clearly shows the symmetry of phase II has to be lower than the previously proposed tetragonal symmetry [28]. We also relaxed all the atoms in the monoclinic phase II structure from the Rietveld refinement, and its band structure does not change, indicating that the structure we solved is stable electronically.

In phase III, all the diffraction peaks can be well indexed and fit by the orthorhombic *Aba2* structure (Supplementary Fig. 1(b)). The clear separation between diffraction peaks like $(220)$ and $(202)$ indicates unequal *b* and *c*, and thus refutes the early proposed tetragonal symmetry [28]. We also performed first-principles calculations to search globally for the stable structure of phase III through USPEX with fixed experimental volume in phase III region [36–38], and found that the *Aba2* structure is a stable phase.

Two different Ag sites, Ag1 and Ag2, generally exist for phases I, II, and III. The evolution of the Ag1 to nearby Te distances as a function of pressure is plotted in Fig. 2. From phase I to phase II, the striking difference is the Ag1-Te2 distance dropped from 3.93 Å to 3.34 Å, increasing the coordination number for Ag1 from four to five, as shown in Fig. 1(b). The coordination of Ag2 keeps to be tetrahedra in all three phases. From phase II to phase III, pressure drives the lower symmetry monoclinic structure into a higher symmetry orthorhombic structure.

Fig. 3 presents the evolution of volume per $Ag_2Te$ formula unit with increasing pressure, and the inset shows normalized lattice parameters for phase III. The volume per unit $Ag_2Te$ formula decreases continuously from phase I to phase III. In phase III, the *c* direction lying in the layers stacking plane is slightly more compressible than *a* and *b* directions. When applying external pressure, the Ag1 pyramids and Ag2 tetrahedra experience increasing distortion to cause unequal compression rates for *a*, *b*, and *c*.

In order to determine the electronic structure of different phases, we carried out first-principles calculations. Phase I is proposed as a topological insulator [4] with topological surface states including an odd number of gapless Dirac cones inside the bulk band gap. Our band structure calculation for phase I agrees well previous study [4], shown in Fig. 4(a). The red dots show the character of the Ag s orbital. The band inversion can be seen clearly near the Γ point. In comparison, shown in Fig. 4(b), phase II becomes a semimetal as there are only a few states crossing the Fermi level, for instance, the valence band between Z and Γ raised above Fermi level. The band inversion around the Γ point still exists, so its topologically non-trivial nature is maintained. The band structure of phase III is shown in Fig. 4(c). The large band overlap between the conduction and valence bands clearly shows that phase III becomes a bulk metal, and thus it is unnecessary to define its topological nature.

Based on these electronic structures, the previous high pressure magneto transport results of on $Ag_2Te$ [27] can now be understood. For phases II and III, different MR responses are expected based on conventional band theory with a closed Fermi surface [40]. In phase II, the bulk metallic properties dominate the surface properties, thus the MR will have more quadratic character. In phase III, the system becomes a bulk metal, and thus results in a completely quadratic MR response. Previously reported low temperature transport measurements on $Ag_{2-\delta}Te$ at 1.01, 1.35, and 1.71 GPa showed decreasing electrical resistivity and gradual weakening of LMR, and at a clear quadratic MR response was observed at 1.71 GPa [27]. These results indicate comparatively large band structure evolution near the Fermi surface for $Ag_{2-\delta}Te$ under pressure, and 1.71 GPa should be close to the transition between the insulating phase I and the semi-metallic phase II at 4.2 K.

Fig. 4(d) shows the calculated DOS of phases I, II, and III. Among them, phase III has the highest DOS. From -1 eV to -0.13 eV below the Fermi level, phase II has a lower DOS than phase I. Yet from -0.13 eV above up to 1.25 eV, the DOS of phase II is much greater than that of phase I. In the case of a minor amount of self doping in $Ag_{2-\delta}Te$ and $Ag_{2+\delta}Te$, the crystal structure should be the same and the band structure features will be maintained, with a mere adjustment of the Fermi level. For silver rich $Ag_{2+\delta}Te$ (*n* type), the Fermi level will shift to higher energy. The increase of the DOS will lead to decreased electrical resistivity with increasing pressure from the phase I to phase II regime. For silver deficient $Ag_{2-\delta}Te$ (*p* type), the situation is very interesting: depending on the doping ratio, the shifting Fermi level may drop into the low energy (-0.13~0 eV) region or even lower energy (< -0.13eV) region. In these two regions, completely different transport behavior is expected. We roughly estimated the value of δ that can shift Fermi level down 0.13 eV is 0.02. The earlier room temperature transport measurements showed that phase II was less conductive than phase I, and phase III was twice as conductive as phase I [28]. The reason why phase II became less conductive than phase I is that the measured sample was δ > 0.02 $Ag_{2-\delta}Te$ and phase II has smaller DOS than phase I at that doping ratio. Phase III is the most conductive for its largest DOS near the Fermi level at all doping ratios.

In summary, the structural transitions of $Ag_2Te$ up to 42.6 GPa at room temperature was studied by *in-situ* synchrotron angle dispersive XRD and first-principles calculations. Three phases were identified at high pressure. Phase II is solved to be a monoclinic structure with the same space group to phase I, and phase III has an orthorhombic structure. Pressure allows us to manipulate the electronic band structure of $Ag_2Te$ and tune its electrical and magneto transport properties dramatically, as $Ag_2Te$ changes from narrow gap insulating phase I to semi-metallic phase II, and then to metallic phase III.

**Acknowledgment:** We thank beamline scientists C. Park, S. Sinogeikin, and Y. Meng in HPCAT, APS for their assistance, and meaningful discussions with W. Jing and Y. He. ZZ is supported through Stanford Institute for Materials and Energy Sciences (SIMES). SW is supported by EFree, an Energy Frontier Research Center funded by the U.S. Department of Energy (DOE), Office of Science, Office of Basic Energy Sciences(BES) under Award Number DE-SG0001057. HZ is supported by the Army Research Office, No. W911NF-09-1-0508.

Figure Captions:

FIG. 1(a). Representative XRD patterns for $Ag_2Te$ under pressure up to 42.6 GPa ($\lambda$=0.37379 Å). Either new peaks or peaks with different intensity are marked by circles, diamonds and asterisks for phases II, III, and IV correspondently. Arrows indicate diffraction peaks from the Ne pressure medium. (b) Schematic view of $Ag_2Te$ phases I, II, and III structures.

FIG. 2. Ag1 to nearby Te distance vs pressure from experiment using Ne. Errors given by GSAS – EXPGUI package are smaller than marker sizes.

FIG. 3. Volume per $Ag_2Te$ formula unit versus pressure, inset shows normalized cell parameters $a/a_0$, $b/b_0$ and $c/c_0$. Filled and half filled circles are from experiments with Ne and silicone oil as pressure medium respectively. Errors given by GSAS – EXPGUI package are smaller than marker sizes.

FIG. 4. Calculated band structure of (a) phase I, (b) phase III, and (c) phase III. The red dots indicate the projection of the Ag s orbital. In (a) and (b), the Ag s orbital goes down to be occupied from unoccupied around $\Gamma$, which indicates the band inversion. Fermi level is shifted to 0 eV. These band structures present (a) the phase I with a narrow band gap, (b) the semi-metallic phase II, and (c) the metallic phase III. (d) Calculated DOS of phases I, II, and III.

Figure 1

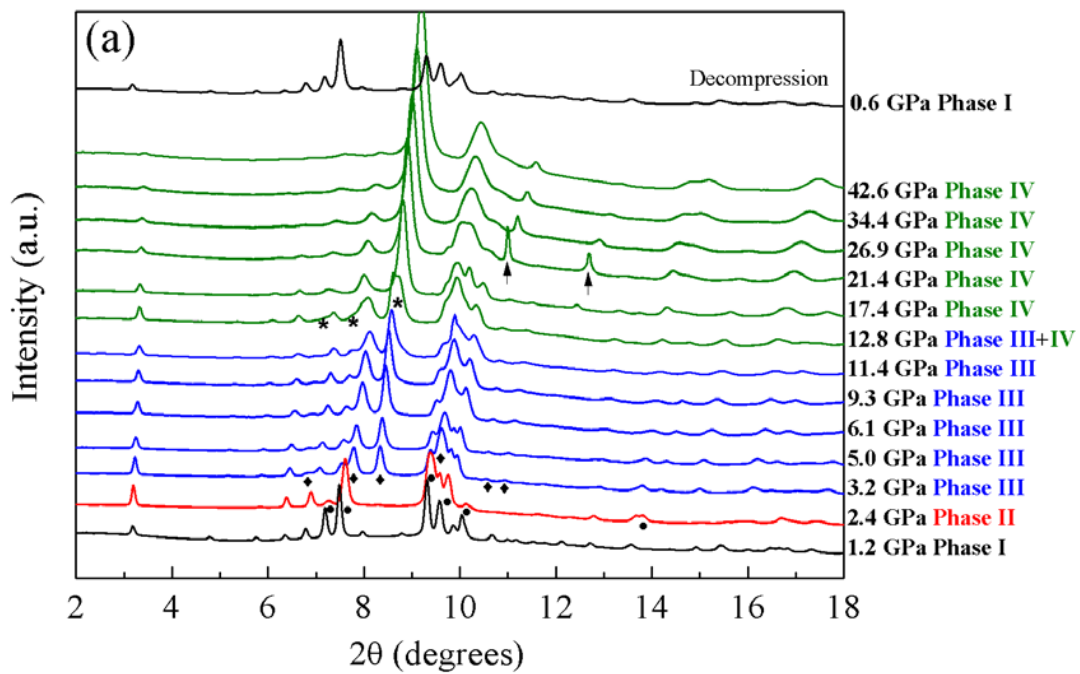

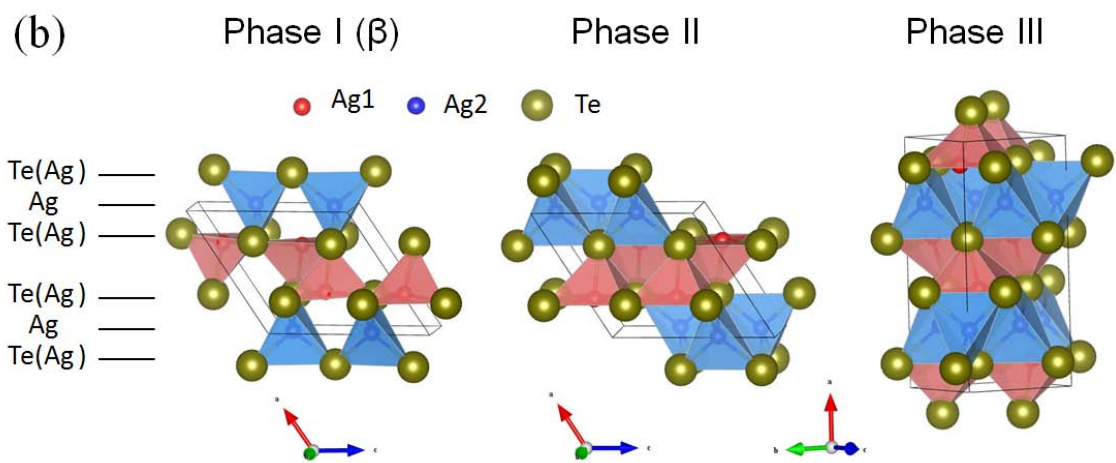

Figure 2

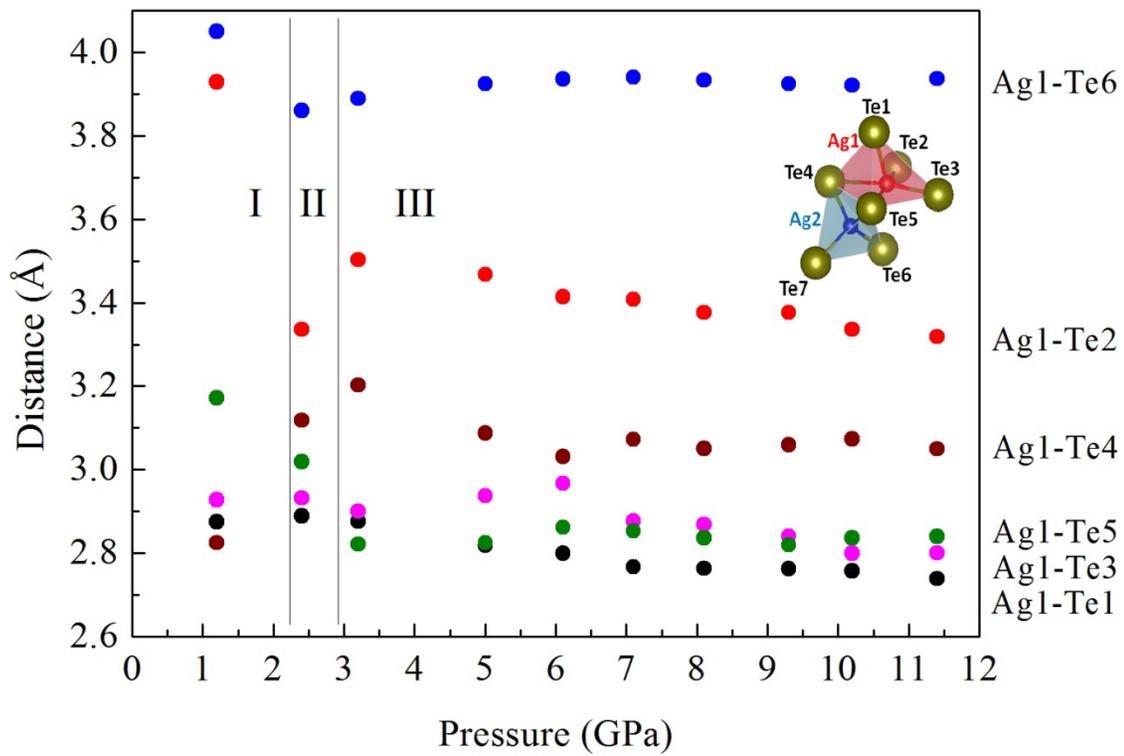

Figure 3

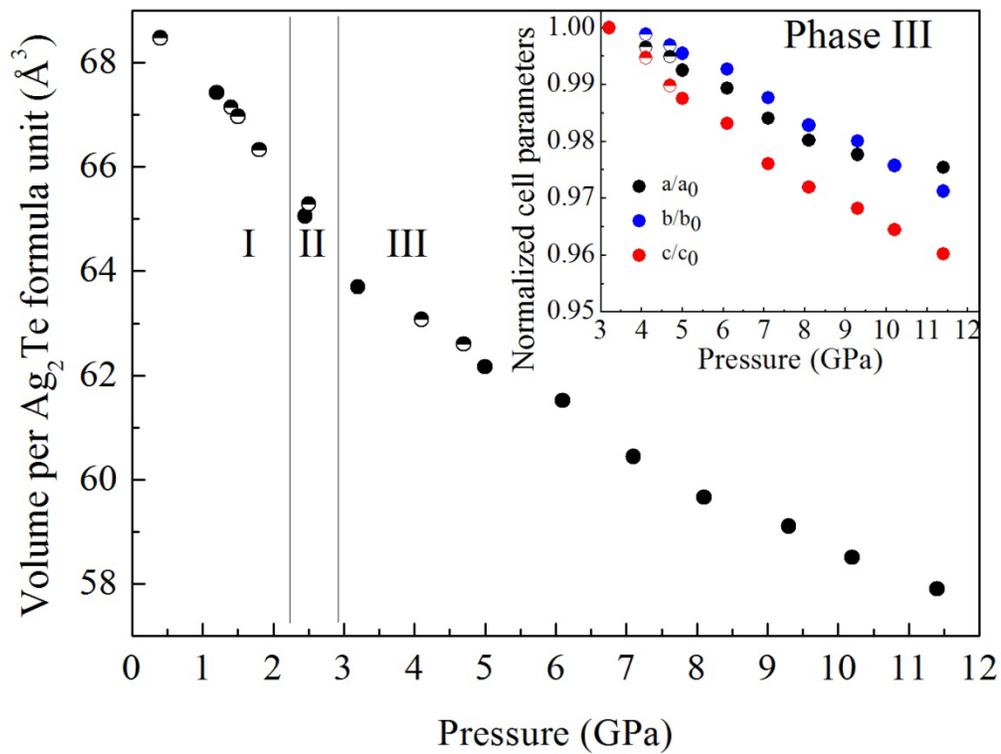

Figure 4

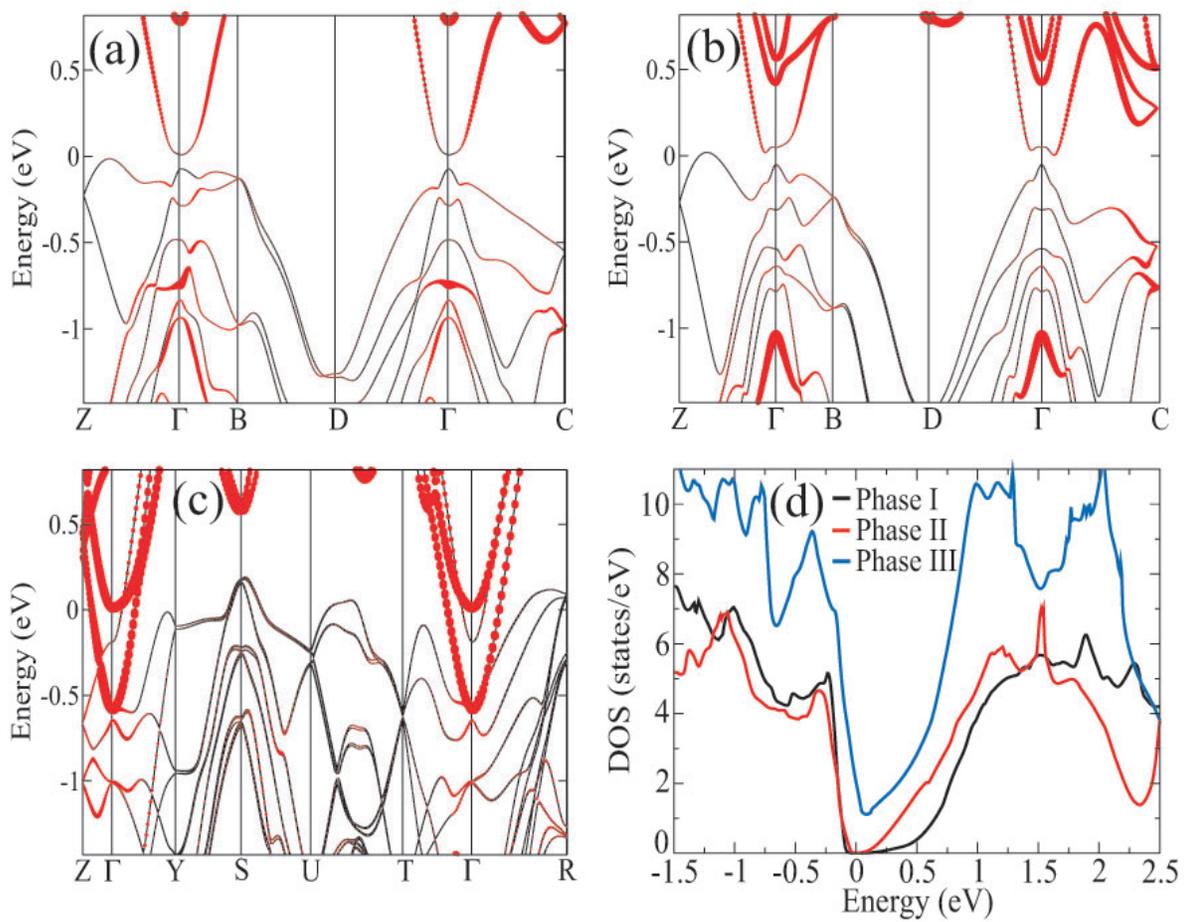

# Pressure induced structural transitions and metallization in Ag$_2$Te

Zhao Zhao[1], Shibing Wang[2,3], Haijun Zhang[1], Wendy L. Mao[2,4]

[1] *Department of Physics, Stanford University, Stanford, CA 94305, USA*

[2] *Department of Geological and Environmental Sciences, Stanford University, Stanford, CA 94305, USA*

[3] *Stanford Synchrotron Radiation Light Source, SLAC National Accelerator Laboratory, Menlo Park, CA 94025, USA*

[4] *Photon Science, SLAC National Accelerator Laboratory, Menlo Park, CA 94025, USA*

## Supplementary materials

**Supplementary TABLE I.** Rietveld refinement results for phases I, II, and III. Errors are given by GSAS-EXPGUI package.

| Pressure (GPa) | Space group | Volume (Å³) | Z | Cell parameter | | | | Position parameter | | | |
|---|---|---|---|---|---|---|---|---|---|---|---|
| | | | | a (Å) | b (Å) | c (Å) | β(°) | Atom | x | y | z |
| Phase I | | | | | | | | Te (4e) | 0.272(2) | 0.362(3) | 0.749(2) |
| 1.2 | P2₁/c (14) | 269.8(2) | 4 | 8.115(1) | 4.465(1) | 8.960(2) | 123.79(1) | Ag1 (4e) | 0.323(4) | 0.810(3) | 0.995(4) |
| | | | | | | | | Ag2 (4e) | -0.003(2) | 0.167(1) | 0.358(2) |
| Phase II | | | | | | | | Te (4e) | 0.285(2) | 0.269(1) | 0.755(2) |
| 2.4 | P2₁/c (14) | 260.6(4) | 4 | 7.972(2) | 4.415(2) | 8.791(2) | 122.60(2) | Ag1 (4e) | 0.289(2) | 0.796(3) | 0.005(2) |
| | | | | | | | | Ag2 (4e) | -0.001(2) | 0.212(2) | 0.392(3) |
| Phase III | | | | | | | | Te (8b) | 0.395(3) | 0.220(4) | 0.678(2) |
| 3.2 | Aba2 (41) | 509.7(5) | 8 | 13.305(3) | 6.317(2) | 6.064(3) | — | Ag1 (8b) | -0.104(2) | 0.834(2) | -0.347(2) |
| | | | | | | | | Ag2 (8b) | 0.240(3) | 0.943(2) | 0.935(2) |

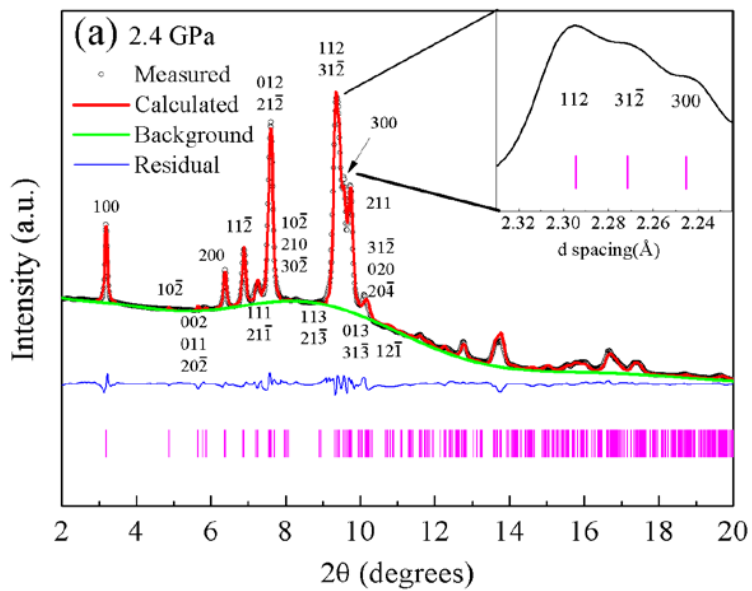

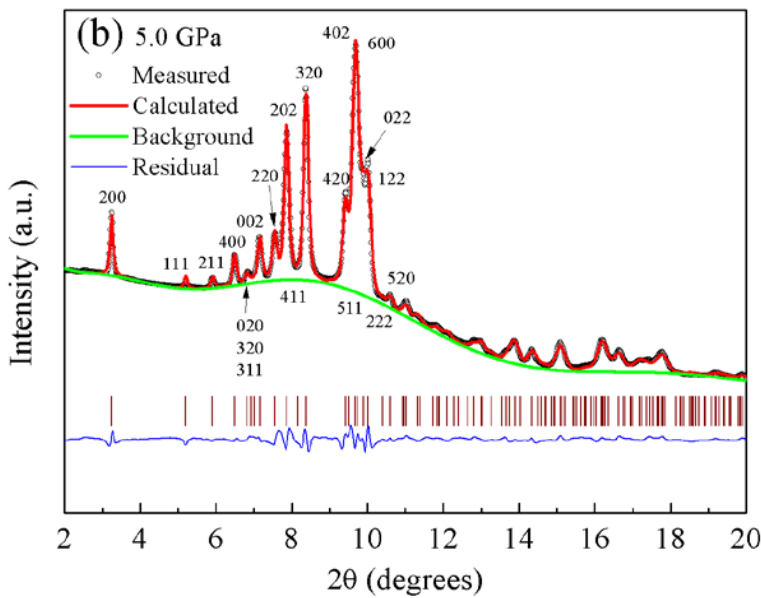

**Supplementary Fig. 1.** Rietveld refinement results for (a) phase II at 2.4 GPa, inset shows the splitting of peaks at 2θ near 9.5° clearly, (b) phase III at 5.0 GPa. The red lines and open circles represent the Rietveld fit and the observed data respectively, and the blue line gives the residual intensities. The vertical bars indicate the predicted peak positions.